\documentclass[10pt,aps,showpacs,floatfix,raggedbottom,floats,superscriptaddress,twocolumn]{revtex4-1}

\pdfoutput=1
\usepackage{graphicx,latexsym}
\usepackage{dcolumn}
\usepackage{amssymb,amsmath,bm}

\usepackage[hidelinks=true,breaklinks=true]{hyperref}

\hypersetup{
  colorlinks   = true, %Colours links instead of ugly boxes
  urlcolor     = black, %Colour for external hyperlinks
  linkcolor    = black, %Colour of internal links
  citecolor    = black %Colour of citations
}

\begin{document}

%-----------------------------------------------------------------
\title{Readout of fluorescence functional signals through highly scattering tissue}

\author{Claudio Moretti}
 \email{claudio.moretti@lkb.ens.fr}
 \affiliation{ 
Laboratoire Kastler Brossel, ENS-Universit\'e PSL, CNRS, Sorbonne Universit\'e, College de France, 24 Rue Lhomond, F-75005 Paris, France
}

\author{Sylvain Gigan}
 \email{sylvain.gigan@lkb.ens.fr}
 \affiliation{ 
Laboratoire Kastler Brossel, ENS-Universit\'e PSL, CNRS, Sorbonne Universit\'e, College de France, 24 Rue Lhomond, F-75005 Paris, France
}

\begin{abstract}
Fluorescence is a powerful mean to probe information processing in the mammalian brain\cite{Helmchen_imaging_2011}. However, neuronal tissues are highly heterogeneous and thus opaque to light. A wide set of non-invasive or invasive techniques for scattered light rejection, optical sectioning or localized excitation, have been developed, but non-invasive optical recording of activity through highly scattering layer beyond the ballistic regime is to date impossible.  Here, we show that functional signals from fluorescent time-varying sources located below an highly scattering tissue can be  retrieved efficiently, by exploiting matrix factorization algorithms to demix this information from low contrast fluorescence speckle patterns.
\end{abstract}

\date{\today}

\maketitle

% introduction paragraph (which is not an abstract; read above)

In the last decades novel light-enabled tools established new paradigms in neuroscience\cite{boyden_millisecond-timescale_2005, chen_ultrasensitive_2013, Helmchen_imaging_2011}, and among them the emergence of fluorescence functional indicators revolutionized the way to monitor information processing through the brain of different animal models, with unprecedented combination of contrast, resolution and specificity\cite{livet_transgenic_2007, weisenburger_guide_2018}.  
With this approach, optical resolution is often not paramount, and in general only a coarse (cell) resolution is needed\cite{prevedel_fast_2016}. Furthermore, when the location of neurons is known, it is possible to avoid slow raster-scanning techniques and image only the needed location at high frame-rate \cite{iyer_fast_2006, katona_fast_2012, grewe_high-speed_2010, prevedel_fast_2016, bovetti_simultaneous_2017}. 

However, brain tissues are usually opaque, and light emitted or delivered at depth in the brain is often quickly subject to multiple scattering events. This results in a loss of directionality after few scattering lengths, corresponding to a few hundred microns, and ultimately means that all wide field or scanning  microscopy techniques fails at depth. While the brains of simple organisms are sufficiently small and/or transparent so they can be imaged in totality, for instance C.Elegans, drosophila or zebrafish\cite{weisenburger_guide_2018}, mammalian brain, starting with its most common animal model, the mouse, is too large and too scattering to image in full. When imaging is performed in superficial layers, it is  possible to implement wide field recording with multi-site multiphoton excitation \cite{nikolenko_slm_2008, bovetti_simultaneous_2017}, or with wide-field excitation and \textit{a-posteriori} demixing, exploiting the few forward scattered or ballistic photons as a seed to separate the individual neuron contributions \cite{zhou_efficient_2018, nobauer_video_2017, pegard_compressive_2016}. However, observing neuronal activity beyond a millimeter in the cortex or through the skull, is to date extremely challenging. In this depth range, in the multiple scattering regime, several techniques have been introduced to focus light and image using wavefront shaping\cite{horstmeyer_guidestar-assisted_2015, rotter_light_2017}. Fluorescence is conventionally considered very incoherent, so these techniques based on coherence do not straightforwardly apply. However, it has been  shown that it is still possible to reconstruct a fluorescent object hidden behind a scattering medium, by analyzing spatial correlation within  a single low contrast fluorescent speckle \cite{katz_non-invasive_2014, hofer_wide_2018, chang_single-shot_2018, stern_non-invasive_2018, xu_imaging_2018}. These techniques require thin media (with the so-called memory effect \cite{judkewitz_translation_2015}) and limited object size\cite{hofer_wide_2018, chang_single-shot_2018, katz_non-invasive_2014}, restricting their application at depth in tissues.

However, precise neuron localization is not always needed to retrieve important insights on the brain function. This is the case for instance in some electrophysiological recording techniques,  where activity of multiple neurons is often captured simultaneously, with only a coarse localization and no cellular sub-population selectivity\cite{hong_novel_2019}. The use of genetically encoded fluorescence markers and optical readout would allow to expand such recording to precise targeting, minimal invasiveness, and single neuron resolution\cite{lin_genetically_2016, livet_transgenic_2007, Helmchen_imaging_2011, weisenburger_guide_2018}.

Here, we therefore choose to completely relax this imaging  constraint: our goal is not to retrieve an image of a fluorescent object, nor localize it, but instead  to retrieve the temporal activity from deeply buried fluctuating sources (ideally the functional activity of a set of neurons), by recording in wide-field  their fluorescence.
We rely on the fact that each source, after scattering through a thick opaque medium, will generate an extended but well defined spatial pattern at the detector, i.e. a speckle, which will be its unique \emph{footprint}. This pattern  will be modulated in time and summed incoherently with the other source's footprints on a camera.
Here, we show experimentally that it is possible to exploit these low contrasted fluctuating speckle patterns from extended fluorescence sources to extract functional signals even through highly scattering tissues, using an advanced signal processing algorithm, based on non-negative low-rank matrix factorization.  We demonstrate the technique through a mouse skull, using large fluorescent sources and realistic temporal signals.

\begin{figure}[hb]
\includegraphics[width=\linewidth]{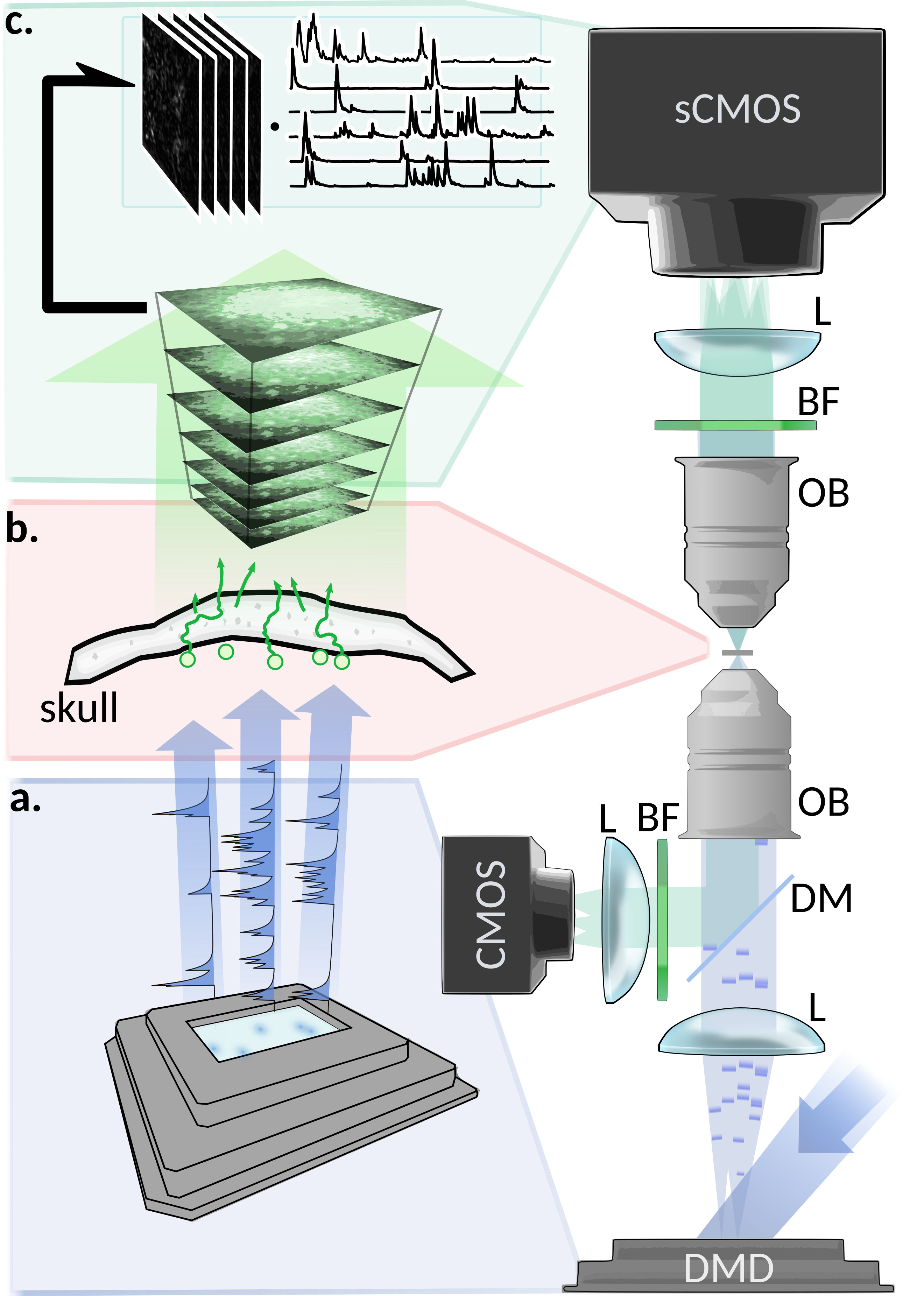}
\caption{\label{fig:fig1}Schematic of hardware setup and operation of the fluorescence activity recording through the highly scattering sample. A blue laser illuminates a DMD in order to excite the fluorescent beads with well defined spatio-temporal patterns (a.). The fluorescence is scattered by a mouse skull (b.), and the resulting speckle is collected by a microscope objective (OB), spectrally filtered (BF), and finally collected on a \mbox{sCMOS} camera (c.). A second camera (CMOS), placed after a dichroic mirror (DM) below the sample, allows the recording of the ground-truth of the induced fluorescent sources activity. An NMF-based algorithm has then been used to extract from the camera recording the temporal activity, together with the speckle pattern associated to each fluorescent source.}
\end{figure}

In order to test this approach, we designed an experimental setup providing a reliable ground truth of the light emitted from the fluorescent object, and the activity which they encode. We simulate the temporal activity of a small network of \emph{synthetic neurons} made by $10\mu m$ fluorescent spheres (close to the size of common neurons bodies\cite{bovetti_simultaneous_2017}), where the emission spectra has been chosen to be close to common green fluorescent activity indicators, and  using publicly available neuronal activity recordings and fluorescence indicator's physiological models\cite{chen_ultrasensitive_2013, deneux_accurate_2016}. As  described in \emph{methods}, the temporal activity is generated by exciting the beads using a blue laser and a spatial light modulator (fig.\ref{fig:fig1}).
The bead's fluorescence emission then experiences the scattering through an \emph{ex-vivo} mouse skull (thickness $\simeq 300\mu m$, and $l_s\simeq 40\mu m$ \cite{soleimanzad_optical_2017}). The time-fluctuating low contrasted speckles are then recorded on a sCMOS camera for the analysis through a $10\times$ water objective and a standard fluorescence emission filter.

% Results
% 

\begin{figure*}
\includegraphics[width=\linewidth]{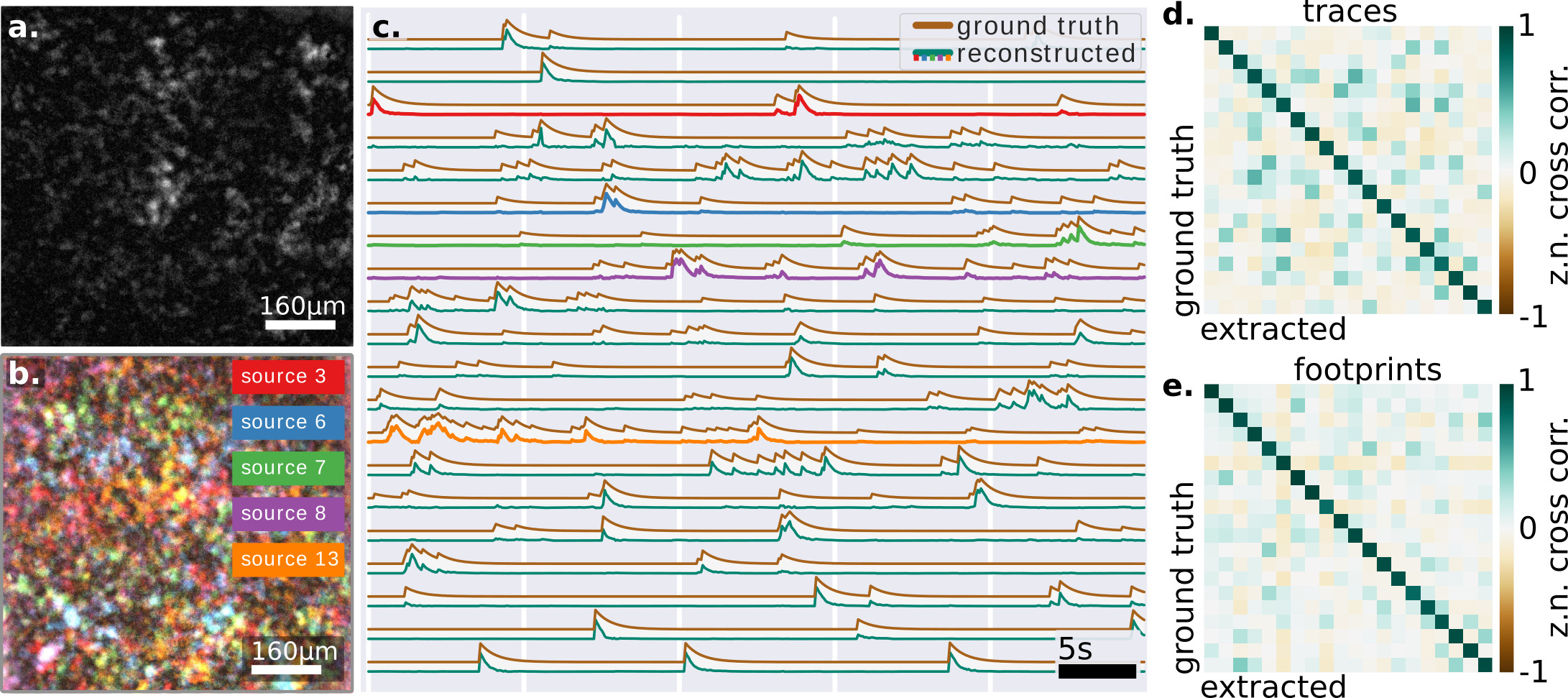}
\caption{\label{fig:fig2}(a.) Filtered speckle pattern at the camera of a single fluorescent emitter. (b.) Superposition of five different filtered footprints. (c.) Visual comparison of extracted traces with their respective ground truth. Traces associated with the footprints in b. are highlighted with their respective colors. (d.) Zero-norm cross-correlations of extracted traces and their sorted ground-truth. (e.) Cross-correlations of extracted footprints and their ground-truth, with the same sorting as d. Zero-norm cross-correlation are on average $0.87 \pm 0.03$ in the diagonal, and $0.02 \pm 0.15$ outside (t-tests $p\leq10^{-4}$).}
\end{figure*}

When a single bead is illuminated, the intensity pattern recorded at the camera shows a clear speckle structure (supp.fig.1a).  The result after  background removal is shown in fig.\ref{fig:fig2}.a and supp.fig.1.c. 
Because of the relatively wide emission spectrum and the large size of the beads, the contrast of the pattern in fig.\ref{fig:fig2}.a is lower than unity, which would correspond to a fully developed speckle from a point-like monochromatic source\cite{goodman_speckle_2007}. In fact, we expect a speckle pattern with a contrast $C=1/\sqrt{N}$, where $N$ represent the number of independent degrees of freedom of the light after going through the scattering medium\cite{goodman_speckle_2007}. These degrees of freedom would result from unpolarized and spatially incoherent emission, since it originates from an extended object, and from independent spectral speckle components.
When considering our real measurements $x(t)$, we can try to retrieve the temporal traces $\tau_i(t)$ and the spatial footprints $\zeta_i$ by finding the solution to the minimization:

\begin{equation}
\min_{\zeta, \tau} \left| x(t) - \sum_{i=1}^{r}\zeta_i\tau_i(t) \right| = \min_{Z, T}\left| X - Z T \right|
\end{equation}

In this form the problem can be formulated as a low-rank factorization, where a matrix $X \in \mathbb{R}^{p\times t}_{>0}$ is approximated with two lower rank matrices $Z\in \mathbb{R}^{p\times r}_{>0}$  (the footprints) and $T\in \mathbb{R}^{r\times t}_{>0}$ (the time traces), where $r$ is the desired rank, $p$ are the pixels, and $t$ the frames. In our case, $r$ corresponds to the number of fluorescent sources which compose the recorded signal (which can be independently inferred from the measurements, see supp.fig.4). Since the extracted signals and their demixed speckle footprints are real-valued, the matrix factorization can take advantage of a \emph{non-negativity} constrains with no loss of generalization. This minimization problem falls in the class of the Non-negative Matrix Factorization (NMF) framework, which has been already implemented with outstanding results in other functional imaging signal extraction techniques \cite{pnevmatikakis_simultaneous_2016, nobauer_video_2017, zhou_efficient_2018}, albeit for shallow depth when performed in wide field. 

The random nature of the speckle generation, the strong independence of the $\zeta_i$ vectors, and the physiologically sparse temporal activity, allow the NMF algorithm to converge to a solution with an initialization based on Singular Value Decomposition (SVD) (supp.fig.5). The retrieved solution shows a remarkable correlation both with the known ground truth of the temporal activities and spatial footprint, which suggest that the retrieved solution is close to the global optimum (fig.\ref{fig:fig2}.c-d). 

\begin{figure}
\includegraphics[width=\linewidth]{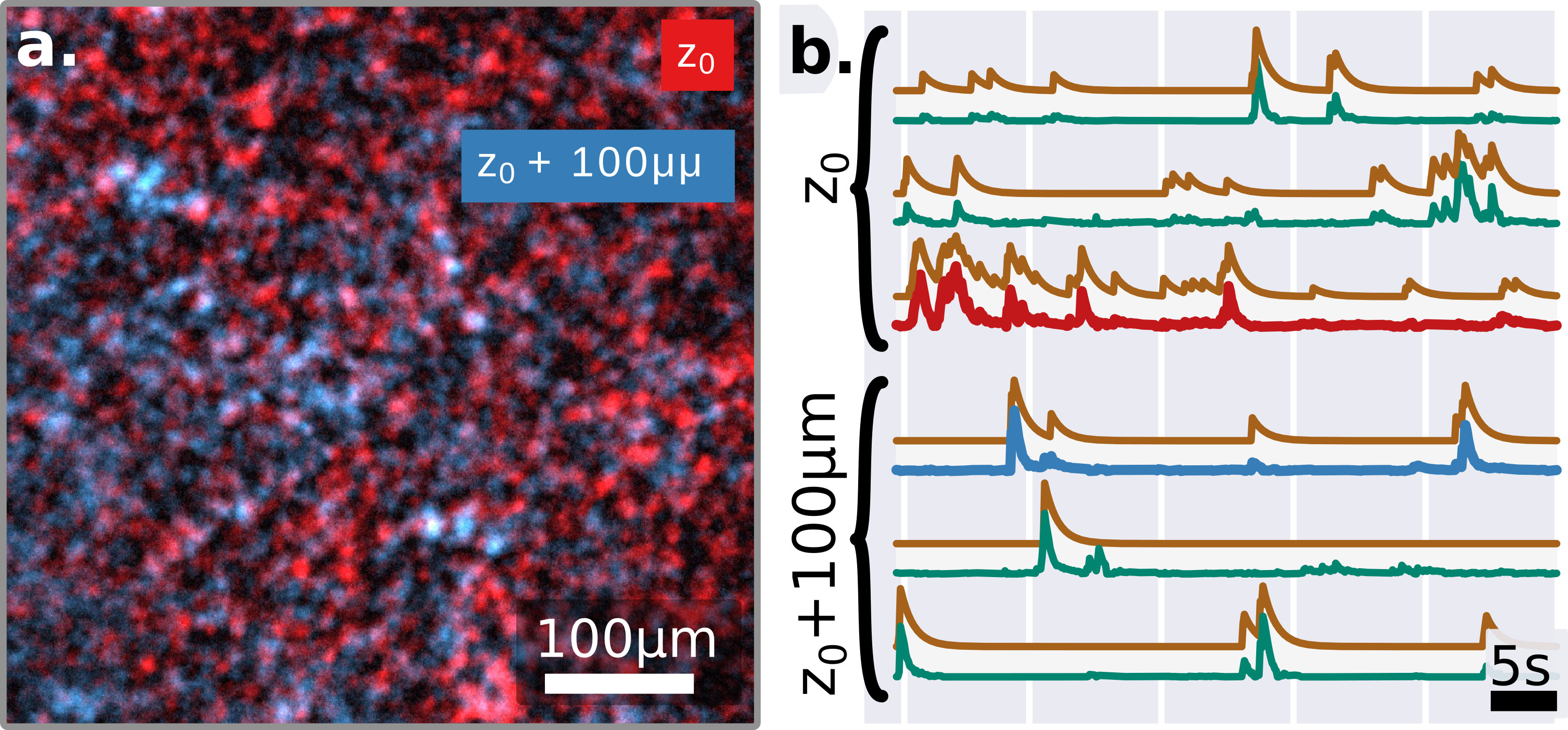}
\caption{\label{fig:axial} Source unmixing when speckles (a.) from two planes were \emph{a-posteriori} incoherently superimposed from two separate recordings of 10 sources at depth $z_0$ and 10 sources at depth $z_0 + 100\mu m$. Representative traces are plot in (b.), where the traces relatives to the footprints of a. are highlighted. Zero-norm cross-correlations of extracted footprints and their ground-truth, with the same sorting as d. Zero-norm cross-correlation are on average $0.90 \pm 0.05$ in the diagonal, and $0.02 \pm 0.11$ outside (t-tests $p\leq10^{-4}$).}
\end{figure}

Since this approach relies only in the randomization of the wave-front performed by the scattering tissue, the relative axial position of the emitter should not compromise the ability to demix  sources located at different depths. This property is demonstrated in the results displayed in fig.\ref{fig:axial}, where are shown few representative traces of two set of 10 sources which have been artificially placed at different depth, $100\mu m$ apart (see methods).

% discussion
%

In both fig.\ref{fig:fig2}.c and fig.\ref{fig:axial}.b we can see the good correlation of the extracted traces with the ground-truths, which is quantified in fig.\ref{fig:fig2}.d-e.
Another important point is the simplicity of the algorithm, which contrarily to implementation at low depth \cite{nobauer_video_2017, pnevmatikakis_simultaneous_2016} does not require an initial guess of the source location (see methods). In practice, we believe the randomization itself (performed by the multiple scattering) actually helps to perform efficiently the factorization, as it is well known in signal processing\cite{mahoney_randomized_2011}. Here, we report on successfully demixing 20 synthetic neurons (limited by the density of fluorescent beads in our sample). However, we believe, based on the SVD decomposition and residual image contrast, that the technique could probably scale to many more neurons, in particular exploiting longer acquisition sequences or adding priors on the temporal structure of the signals. Another important future direction will be to study whether one could use the footprints to extract information about the shape or localization of the sources.  

% Conclusions
%
In conclusion, we have shown that it is possible to retrieve the individual temporal traces of buried fluorescent sources, despite --and actually taking advantage of-- the random scattering process performed by the tissue itself. Using  realistic parameters for neuroscience samples, we demonstrate that it is possible to exploit low contrasted speckles, resulting from  broadband and spatially extended fluorescence emission, to extract functional signals with a NMF-based algorithm, without the need of an initial guess, nor of any assumption on the temporal signal structure or spatial localization of the sources. Importantly, this technique does not rely on ballistic light, or in the presence of speckle correlation, thus is intrinsically adapted to highly scattering regimes. 
 There are obviously many challenges to overcome to apply this technique in \emph{in-vivo} situations (as background fluorescence, high number of sources, and sample movement), still, our work opens a unique avenue towards wide field functional imaging  in highly scattering mediums at unprecedented depths.

\section*{Acknowledgments}
The authors thank Laurent Bourdieu for providing the biological samples and for numerous discussions, Alipasha Vaziri and Tobias N\"obauer for useful suggestions, Saroch Leedumrongwatthanakun, Jonathan Dong and Antoine Boniface for constructive comments.

%\section*{Author contributions}

% \section*{Competing interests}

\section*{Additional information}
This work was funded by  H2020 European Research Council (ERC) (SMARTIES–724473). S.G. is a member of the Institut Universitaire de France. 

\bibliography{morettietal2019}

\newpage

\section*{Methods}

\subsection*{Setup}

Excitation light is produced by a 473nm DPSS laser (LSR-0473-PFM-00100-01, Laserglow Technologies, CA), expanded through a  to fit a Digital Micromirror Device (DMD; DLP LightCrafter 6500, Texas Instruments, US-TX). Light is then focused into the sample with a $200mm$ lens (LA1708-A, Thorlabs) and a 20x (nominal) objective (Plan-NEOFLUAR 20x 0.5NA, Zeiss, DE; bottom \textbf{OB} in fig.\ref{fig:fig1}). A control optical path is used to collect ballistic light with a dichroic mirror (FF496-SDi01, Semrock, US-NY), a filter (MF525-39, Thorlabs), a $100mm$ tube lens (AC254-100-A, Thorlabs), and a CMOS camera (ACE2014-55um, Basler, DE). The sample is composed by $10\mu m$ beads (FluoSpheres F13081, Thermofisher scientific, US-MA), between two \#1.5 coverslips. Beads emission spectra has been filtered to match common green emitting reporters, as GCaMP. Mouse skull is obtained from young adult mice. Light is collected with a 10x (nominal) objective (UMPlanFl 10x 0.3NA, Olympus, FR; top \textbf{OB} in fig.\ref{fig:fig1}), and is filtered either a $43nm$ FWHM barrier filter (MF530-43, Thorlabs; \textbf{BF} in fig.\ref{fig:fig1}) in most of the experiments, or a $10nm$ FWHM filter (FL05532-10, Thorlabs) if explicitly stated. A $200mm$ tube lens (AC254-200-A, Thorlabs) is then used to create the image at the sCMOS sensor of the main camera (EDGE 2.4, PCO, DE). An epi-illumination pathway has been set as well to support alignment procedures (supplementary figure 6).

\subsection*{Activity Traces}

Synthetic fluorescence traces has been generated using spike activity from real available datasets acquired in mouse visual cortex\cite{chen_ultrasensitive_2013}, and calcium traces have been generated from these spikes using a GCaMP6s physiological model \cite{deneux_accurate_2016}, and resampled at 10Hz. Pulse-width modulation was used to encode light intensity levels with the binary DMD amplitude modulation. A calibration routine has been written in Matlab (MathWorks, US-MA) to precisely match DMD pixels to the sample plane, and so the fluorescent beads positions. Acquisition, DMD control, and their synchronization is performed using Matlab (see script repositories for detailed implementation).

\subsection*{Data Acquisition}

Imaging objective (upper \textbf{OB} in fig.\ref{fig:fig1}) has been focused $2mm$ from the skull surface, to obtain a proper sampling of the pattern. Every recording, the pattern produced by the distinct beads has been recorded one-by-one (fig.\ref{fig:fig2}.b), which provides the speckle footprint ground truth, and allows studying the correlation with the extracted footprints and the cross-correlations among them to estimate the memory effect (supp.fig.2). Exposure time has been set to 400ms. Frames has been cropped to the central zone of the background envelop. For figure\ref{fig:axial}, we recorded both planes sequentially, then \emph{a-posteriori} overlaid the recording, to emulate incoherent superposition of all neurons footprints. All the cross-correlation between traces and footprints has been evaluated as zero normalized cross-correlation to manage the transmission disomogenheities across the sample.

\subsection*{Algorithm}

A proper spatial filter has been shown to be of crucial point to obtain a reliable demixing in the later analysis steps.  A gaussian filter in the Fourier domain is used to remove structures with frequencies higher than the speckles grain size, assumed to be detection noise, and to remove the background envelope. The dataset has been previously sub-sampled in space reaching a shape of ~100x100 pixels, then a singular value decomposition of the dataset has been performed. The resulting eigenvalue distribution has been clusterized using a k-mean algorithm, and the number of sources has been identified (supplementary figure 4).  The factorization has been performed with \emph{decomposition/NMF} class from the \emph{Scikit Learn} Python libray. We didn't use any sparsity constraints nor priors to retrieve the matrices. Initialization has been performed using a NNDSVD algorithm\cite{boutsidis_svd_2008}. Factorization algorithm has been iterated 100 times, but faster convergence was achieved for most of the datasets (supplementary figure 5).

\section*{Data Availability}
\begin{flushleft}
Full datasets are avaiable upon request.\\
Analysis scripts are available at:
\href{https://github.com/m0ro/SpeckledNeurons\_analysis}{github.com/m0ro/SpeckledNeurons\_analysis}\\
Hardware control scripts are available at: \href{https://github.com/m0ro/SpeckledNeurons\_control}{github.com/m0ro/SpeckledNeurons\_control}
\end{flushleft}

\pagebreak
\onecolumngrid
\begin{center}
\textbf{\large Supplementary informations:\\Readout of fluorescence functional signals through highly scattering tissue}\\[1cm]
\end{center}

\setcounter{equation}{0}
\setcounter{figure}{0}
\setcounter{table}{0}
\setcounter{page}{1}
\renewcommand{\theequation}{S\arabic{equation}}
\renewcommand{\thefigure}{S\arabic{figure}}

\begin{figure}[h]
\includegraphics[width=\linewidth]{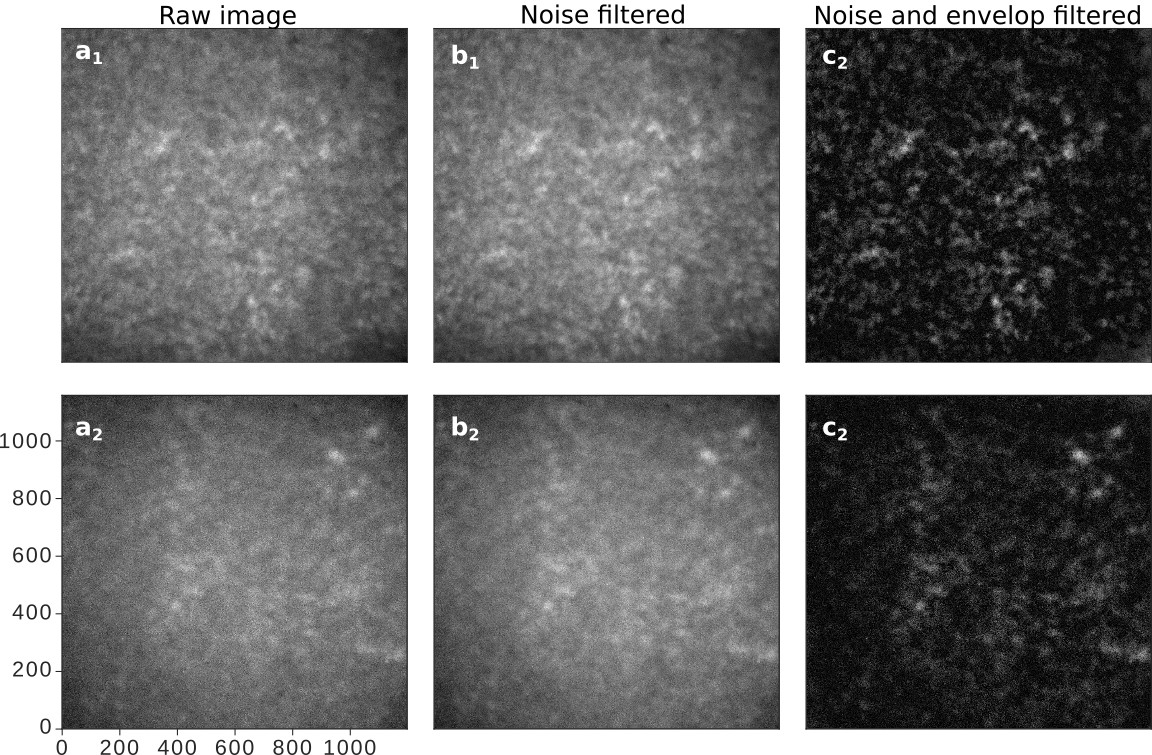}
\caption{\label{fig:supp.filt} Speckle patterns recorded from the sCMOS camera. $\mathbf{a_1}$ shows the un-processed footprint of a single source, while $\mathbf{a_2}$ shows the pattern due to the superposition of 20 sources during an activity recording (the sources have different relative intensity). $\mathbf{b_{1,2}}$ same frames as in the left, filtered with a low-pass Gaussian filter. $\mathbf{c_{1,2}}$ shows the same frames filtered with both low- and high-pass Gaussian filters. Axis units are in camera pixels.} 
\end{figure}

\begin{figure}[h]
\includegraphics[width=0.95\linewidth]{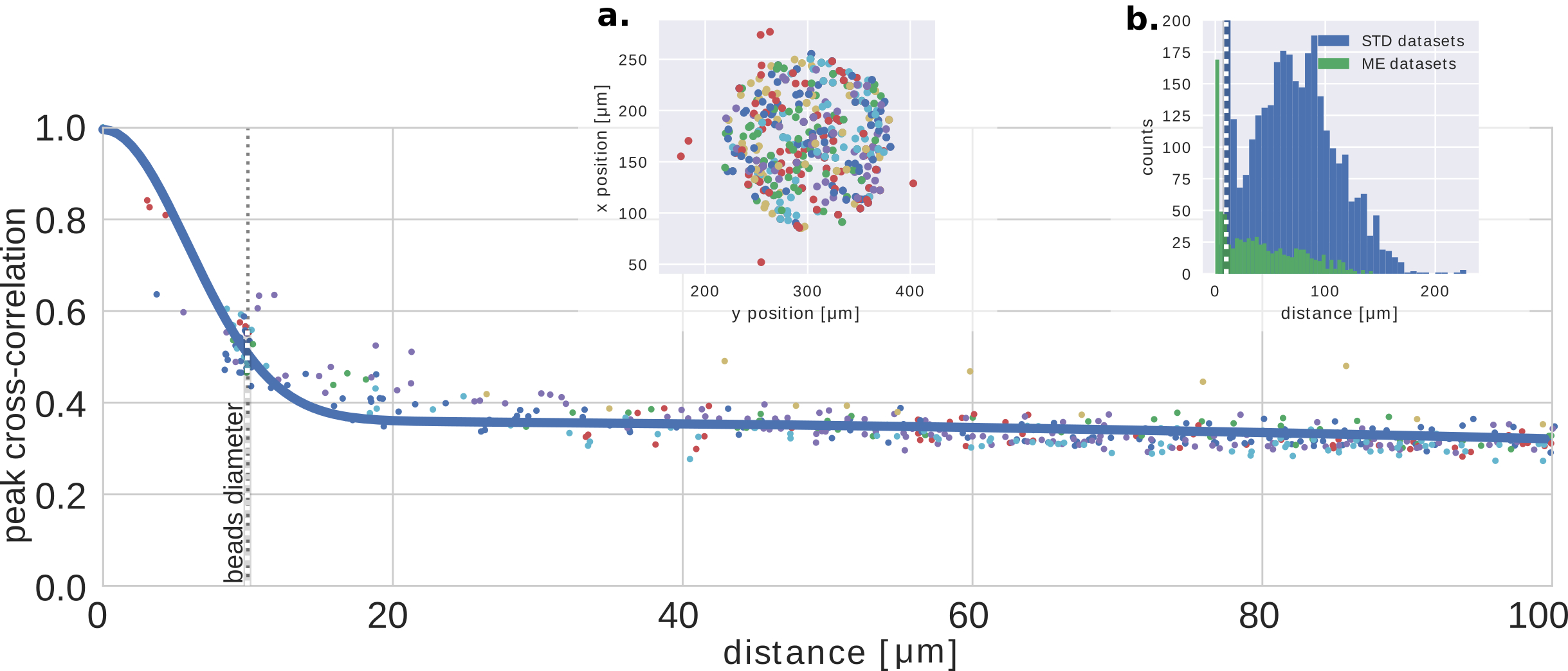}
\caption{\label{fig:supp.me} Plot of the peaks of cross-correlations between different patterns $\zeta_{i,j}$ due to different fluorescent sources located across the samples, over the relative distance $(i,j)$. In the onset \textbf{a.} the positions of the beads, where the colorcode stand for the different data sets. In the onset \textbf{b.} is reported the distance distribution for two different data sets: in green the data set used to estimate the memory effect (ME), while in blue the one used for the activity demixing (STD). The dotted line in the main plot and in the panel b. represent the average diameter of the beads.}
\end{figure}

\begin{figure}[h]
\includegraphics[width=0.5\linewidth]{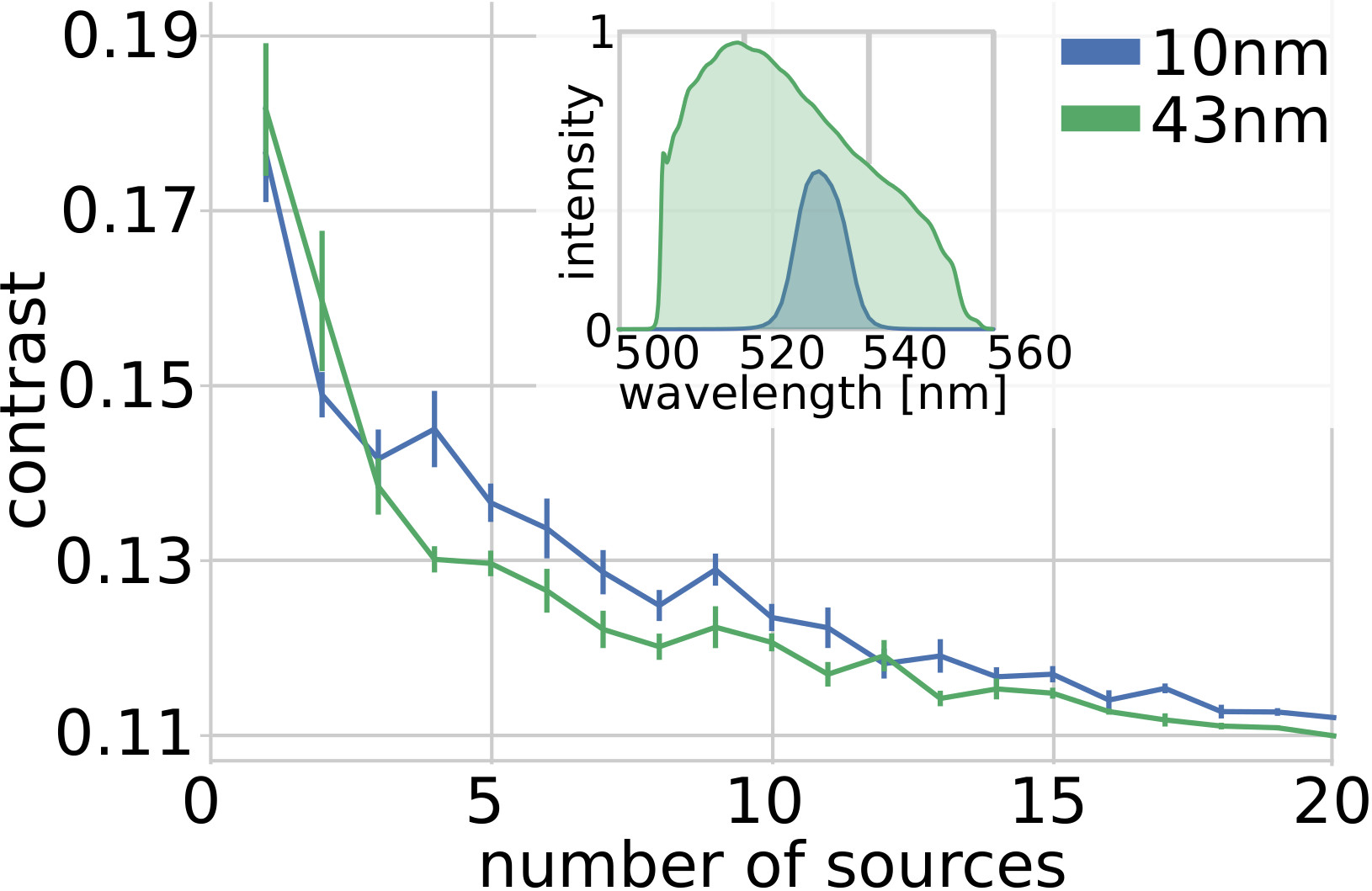}
\caption{\label{fig:supp.spectra} Contrast on the speckle pattern due to the emission from 1 to 20 simultaneously active fluorescence source. Data are shown for both $43nm$ and $10nm$ bandwith barrier filters. In the onset are shown the spectra which results from the expected fluorescent emission filtered by two used filters.}
\end{figure}

\begin{figure}[h]
\includegraphics[width=0.9\linewidth]{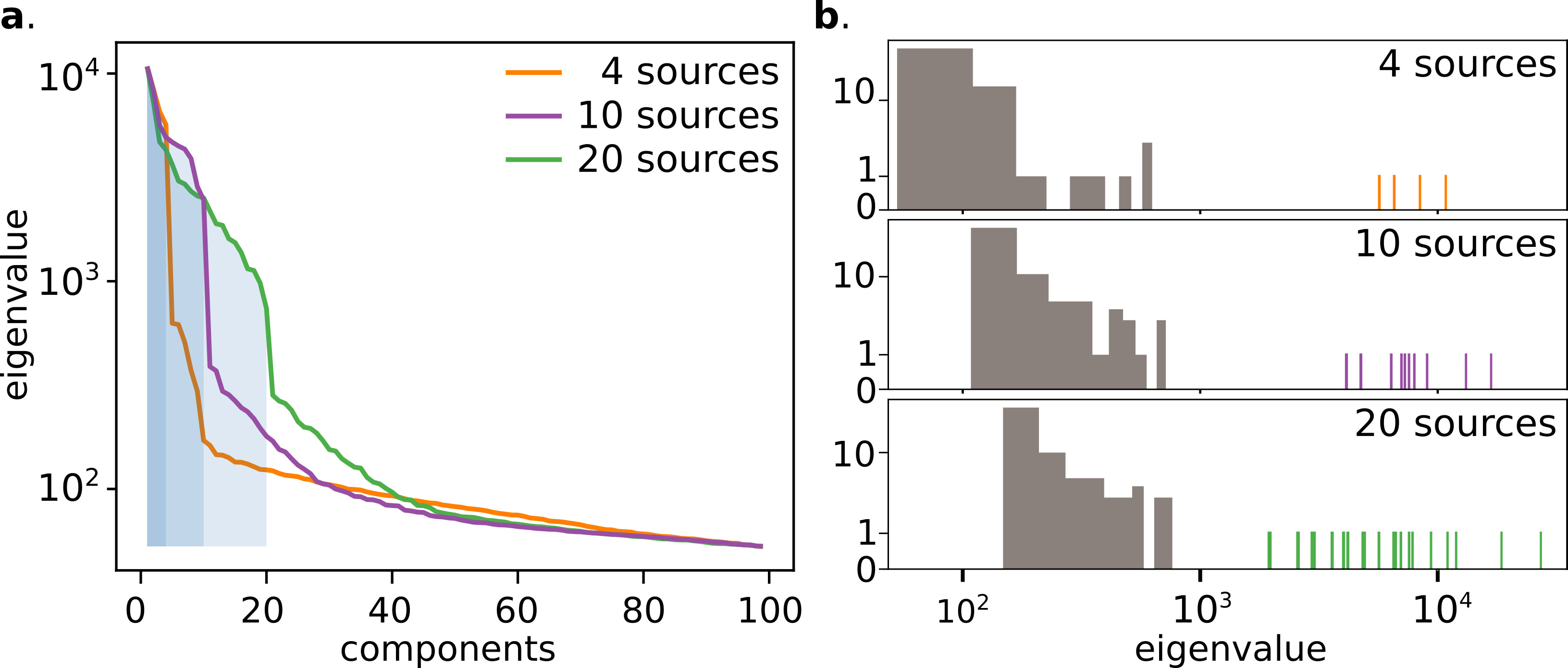}
\caption{\label{fig:supp.svd}Source number identification procedure using single value decomposition and a clustering algorithm. On \textbf{a} the plot of the eigenvalues for 3 sample data sets, with 4, 10 and 20 active fluorescent sources. In \textbf{b} the histogram of the eigenvalues, where highlighted the components which has been separated with a k-mean clustering. Resulting clusters are reported back in \textbf{a} as light blue areas.}
\end{figure}

\begin{figure}[h]
\includegraphics[width=0.7\linewidth]{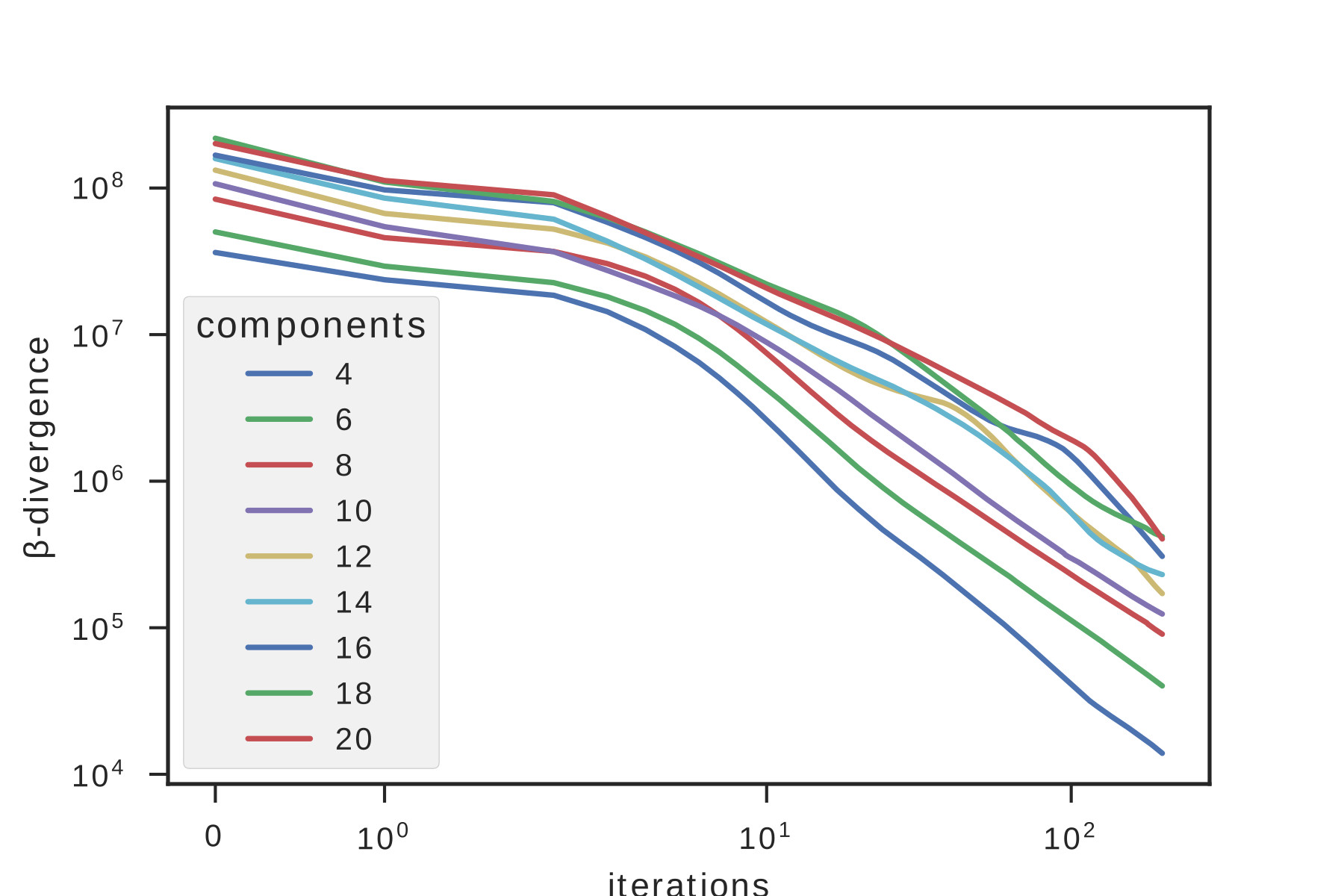}
\caption{\label{fig:supp.axial}Convergence rate of the factorization algorithm, where a different number of emitting sources were present in the sample. The curves report the $\beta$-divergence calculated at every iteration of the NMF algorithm.}
\end{figure}

\begin{figure}[h]
\includegraphics[width=0.5\linewidth]{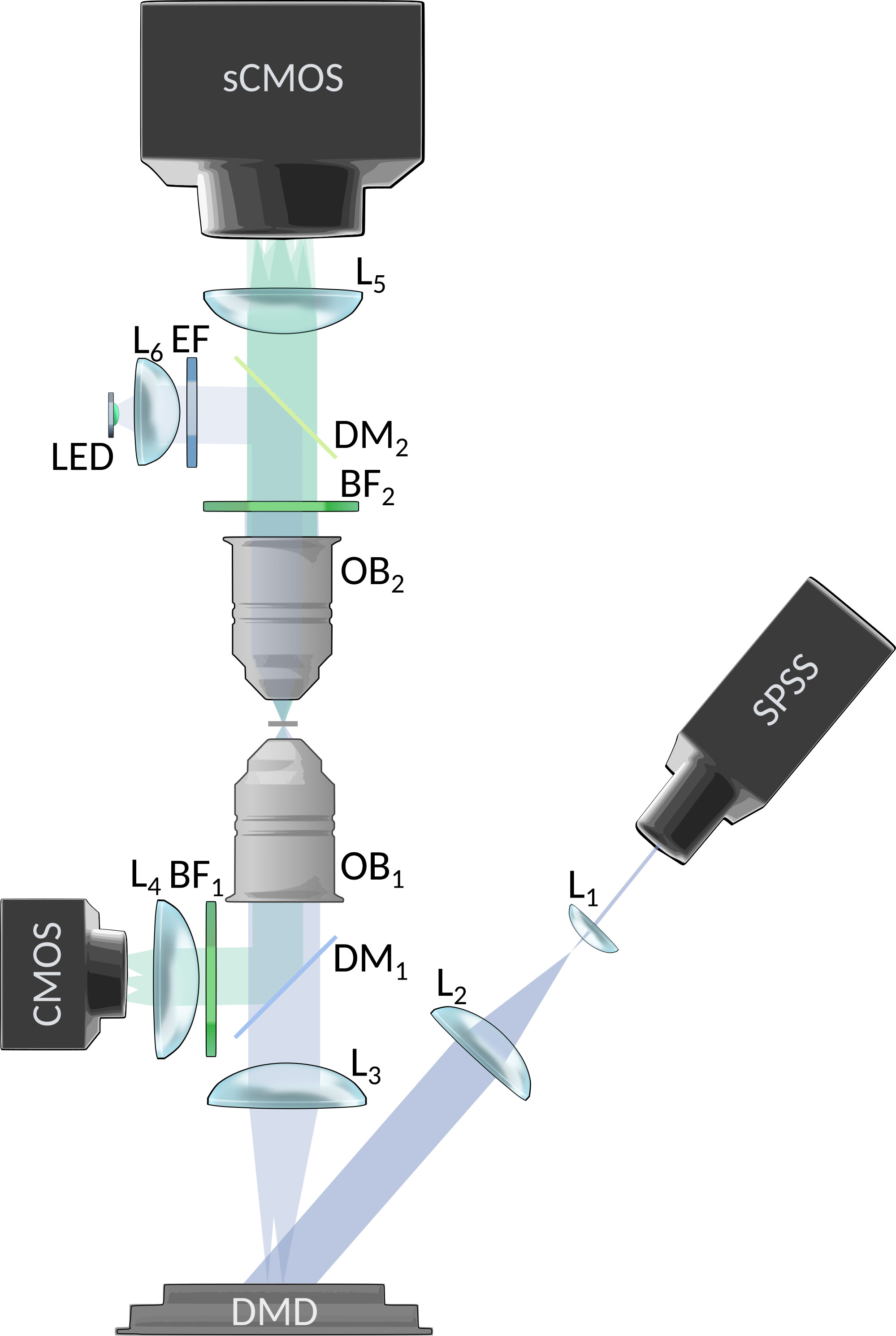}
\caption{\label{fig:supp.axial}
Complete scheme of hardware setup design. The 473nm DPSS laser source is expanded with two singlet lenses (L$_{1-2}$) to properly full-fill the $8mm \times 14mm$ DMD's active window. The DMD's surface is then imaged to the beads layer using a tube lens (L$_3$) and the lower objective (OB$_1$). Proper illumination and calibration procedure are obtained with the information retrieved from the camera CMOS, though a dichroic mirror (DM$_1$), a fluorescence barrier filter (BF$_1$), and a tube lens (L$_4$). Fluorescence emitted from the sample is collected by a microscope objective (OB$_2$), spectrally filtered (BF$_2$), and finally imaged on a sCMOS camera through $L_5$. An upright epi-fluorescence excitation pathway composed by a 470nm LED (ML470L3, Thorlabs), a condenser lens (L$_6$; ACL25416U-A, Tholabs), an emission filter (EF; MF469-35, Thorlabs) and a dichroic mirror (DM$_2$; MD498 Thorlabs), is used to inspect the sample upon necessity.}
\end{figure}

\end{document}